\documentclass{article}
\pdfoutput=1 
\usepackage[utf8]{inputenc}
\usepackage{xcolor}
\usepackage{natbib}
\usepackage{booktabs}
\usepackage{setspace}
\usepackage{geometry}
\usepackage{graphicx}
\usepackage{hyperref}
\usepackage{pythonhighlight}
\usepackage{comment}
\usepackage{tikz}
\usepackage{standalone}
\usepackage{array}
\usepackage{adjustbox}
\usepackage{svg}
\usepackage{changepage}
\usepackage{dutchcal}
\usepackage{caption}
\usepackage{subcaption}
\usepackage{soul}
\usepackage{float}
\usepackage{fancyhdr}

\usepackage[nottoc]{tocbibind}
\usetikzlibrary{mindmap,trees}

\newcolumntype{R}[2]{%
    >{\adjustbox{angle=#1,lap=\width-(#2)}\bgroup}%
    l%
    <{\egroup}%
}
\newcommand*\rot{\multicolumn{1}{R{75}{1em}}}

\renewcommand{\arraystretch}{1.5}

\title{\textit{Novelpy}: A \textit{Python} package to measure novelty and disruptiveness of  bibliometric and patent data}
\author{Pierre Pelletier and Kevin Wirtz \footnote{Email: \texttt{p.pelletier@unistra.fr}; \texttt{kevin.wirtz@unistra.fr}} \\ \\ \textit{Bureau d'Économie Théorique et Appliquée} \\ \textit{ Universit\'{e} de Strasbourg, Universit\'{e} de Lorraine, CNRS}}
\date{}

\begin{document}
\onehalfspacing

\maketitle

\begin{abstract}
   \textit{Novelpy} (v1.2) is an open-source \textit{Python} package designed to compute bibliometrics indicators. The package aims to provide a tool to the scientometrics community that centralizes different measures of novelty and disruptiveness, enables their comparison and fosters reproducibility. This paper offers a comprehensive review of the different indicators available in \textit{Novelpy} by formally describing these measures (both mathematically and graphically) and presenting their benefits and limitations. We then compare the different measures on a random sample of 1.5M articles drawn from Pubmed Knowledge Graph to demonstrate the module's capabilities. We encourage anyone interested to participate in the development of future versions.
\end{abstract}

\thispagestyle{empty}
\tableofcontents

\thispagestyle{empty}
\newpage 

\section{Introduction}
\label{sec:introduction}

Identifying and tracking relevant pieces of knowledge is still a core issue of Science of science research. A better understanding of knowledge flow dynamics, mechanisms behind the emergence of new ideas, and identification of novel or impactful documents are crucial to foster effective science that will, in turn, help to address future societal challenges \citep{fortunato2018science, foster2021surprise, ocde}.
This article proposes integrating several bibliometrics indicators within a \textit{Python} package. It gathers within a single module novelty or, more broadly, creativity measurements through combinatorial novelty indicators \citep{uzzi2013atypical,foster2015tradition,lee2015creativity,wang2017bias,shibayama2021measuring} and several impact measures, with disruptiveness metrics \citep{wu2019large,wu2019solo,wu_wu_2019,bu2019multi,bornmann1911disruption}.\\

This module is intended for any researcher in the emerging and multidisciplinary field of Science of Science. There is an increasing tendency to create new scientometrics indicators, but there are fewer initiatives to design reproducible experiments. For novelty indicators, there is little to no reference to prior approaches when creating a new indicator; thus, the flexibility in the choice of measures raises the temptation to choose the measure that produces the intended outcome \citep{foster2021surprise}. Only a few studies try to put together a conceptual background of creativity and the formalization of the indicators \cite{foster2021surprise}. This article provides a mathematical and graphical description of these indicators. To the best of our knowledge, it is the first tool that allows computing these metrics. \\

Two macro types of analysis can describe Scientometrics, performance and Science Mapping Analysis (SMA)\citep{moral2020software}. The performance analysis aims to assess scientific actors' activities and their impact. Its purpose is to assign a value to the productivity and pervasiveness of the research conducted by a unit (article, author, institution). SMA ``is mostly directed at monitoring a scientific field to determine its (cognitive) structure, its evolution, and main actors within''\citep{noyons1999integrating}. It takes a snapshot of a part of the scientific system at a given moment to analyze its structure. The present package allows to do performance analysis through disruptiveness measures; it also assesses the creative potential of papers using novelty indicators. Both metrics require science mapping analysis to be measured since they are generated through maps of the structure of science. Inputs, outputs and impacts of these scientific activities are the three perspectives used in bibliometrics analysis \citep{sugimoto2018measuring}\footnote{Input refers to human and financial resources and captures the different interactions of agents in the system at different levels (authors/institutional/country levels). Output results from the research process, the different entities that characterize a document. Finally, impact measures knowledge dissemination generated by an article through citations, attention by the general public or re-utilization of document’s component.}. Entities involved in most combinatorial novelty indicators use only the output part of documents to compute their measures \citep{uzzi2013atypical, foster2015tradition, lee2015creativity, shibayama2021measuring}, except for \cite{wang2017bias} that uses references from future articles to control for re-utilization. Disruptiveness indicators \citep{wu2019large, bu2019multi, bornmann2020disruptive} take the outputs and impacts of a given document to construct their metrics. They are based on both the references and citations of a given document. This module focuses on metrics using outputs (references/ keywords) and impact features (citation/ references and keywords from future articles).\\

While citation is an invaluable source of information, several limitations exist for using the sheer number of citations to evaluate impact. Inter-field (and even intra-field) comparison could be challenging; the sheer number of scientists and the way science is performed is highly variable depending on the research domain (methodology/ solo author vs team publication/ citation habits). The gap in the number of citations is mainly due to the field's structure and does not necessarily represent the documents' quality. This phenomenon becomes an issue when raw numbers are used to measure the importance of research \citep{purkayastha2019comparison}. The same problem arises with self-citation, comparing national and international journals or documents' languages \citep{van2001language}.

Network effects have been observed in citation dynamics, \cite{wallace2012small} showed that scholars tend to cite researchers with whom they have a deeper social connection. They also found that researchers are more likely to cite collaborators of collaborators, hence creating a citation continuum. Articles with international collaborations are more cited due to network effect \citep{wagner2019international}. Other negative citation behaviours arise in \cite{bornmann2008citation}; scholars tend to cite papers to satisfy editors and reviewers, showing an apparent disconnection between citation and actual importance during the creation process. Field-specific issues can be solved using a normalization method or different counting methods of citations (see \cite{waltman2016review} for a comprehensive review). 
One family of normalized indicators is disruptiveness \citep{wu2019large,wu2019solo,wu_wu_2019,bu2019multi,bornmann1911disruption}. These measures analyze how a focal article acts as a bottleneck between future papers and the references of the focal papers. It captures if a document consolidates a domain (i.e. future papers rely on the same pieces of knowledge as the focal paper) or constitutes a starting point for documents coming from various areas (i.e. future paper only uses information from the document). \\

Scientific advance is the result of the individuals' creativeness, where \textit{creativity} is defined as "\textit{held to involve the production of high-quality, original, and elegant solutions to complex, novel, ill-defined, or poorly structured problems}" \citep{hemlin2013leadership}.
Scholars have proposed measurements to complete these impact indicators with creativity indicators, usually called ``atypicality'', ``originality'' or ``novelty'' indicators. The need for quantifying novelty comes from its position as an essential component of the structure of the scientific and economic system. Novelty is at the origin of peer recognition which acts as a "reward system" for individuals. The "priority rule" grants recognition to the first person making the discovery \citep{merton1957priorities, carayol2019}. Novelty is also at the core of the theory developed in evolutionary economics in which technological progress and creativity influence the cyclical nature of the economy \citep{schumpeter1939business,nelson1985evolutionary,amendola2014novelty}. Scientific progress remains elusive, and novelty indicators are intended to approach creativity, as making relevant novel combinations is perceived as innovative \citep{burt2004structural,rodriguez2016research,bornmann2019do}. Earliest novelty indicators focused mainly on past information (i.e. using an entity that has been created the same year) or on the distance between articles from a given year, based on their references' overlapping \citep{dahlin2005invention}.

More recently, scholars have integrated the conceptual framework of knowledge recombination (a combination of pre-existing ideas that leads to invention) into novelty indicators. This concept was already developed by \cite{poincare1910mathematical}. Although he refers to the specific case of science, it can be extended to any type of non-scientific creative process where combinations can be both material and conceptual \citep{winter1982evolutionary}. \cite{weitzman1998recombinant} discussed how knowledge could be generated through a combinatorial process of past ideas and how this can generate economic growth as long as potential new ideas are exploitable. At the same time, an invention does not necessarily arise from combining two components together for the first time. Indeed, it can also arise from creating a new relationship between two already linked components \citep{schumpeter1939business, henderson1990architectural}. It deepens the idea brought by \cite{jacob1977evolution} that scientific advance emerges from looking at something from a new angle rather than incorporating a new instrument. Scientists have proposed a more probability-centred approach to capture this combinatorial process. Instead of focusing solemnly on the degree of novelty of a combination, they look at how unlikely this combination is to happen. The more distant the items in the combination, the more complex and unlikely it is to make this combination. The combination is, therefore, more novel. To solve mathematical problems, Poincaré used the knowledge he found in another field \citep{poincare1910mathematical}. The more distant the fields were, the more insight he gained. Novel documents exhibit higher variance in citation performance. Academics adopting an exploration strategy face a higher risk of failure \citep{fleming2001,foster2015tradition,wang2017bias, ocde}. Scientific documents that have a fair mix of novel and conventional ideas are more likely to be ``sleeping beauties'' than other documents (see \cite{ke2015defining} and\cite{wang2017bias}). The idea of \cite{march1991exploration} that organizations which explore and consolidate existing process/technology are more likely to survive can also be applied in the scientific realm\footnote{Here survival can be expressed as a high citation count.}. Novelty indicators can be applied to different entities (patent, paper, webpage, etc.) using a different unit of knowledge (references, keywords, meshterms, text, others).\\

Most of the packages available in \textit{R} and \textit{Python} deal with performance or SMA. \cite{moral2020software} carried out a detailed and up-to-date review of the different tools and libraries that help researchers in their daily work. Although much work has been done to study citation, co-authorship or any coupling, novelty and disruptiveness indicators are still unavailable, and researchers have to code these metrics themselves. Concerning the reproducibility of novelty studies, only \cite{shibayama2021measuring} shared their code on Github to calculate their new novelty indicator, but this is still an isolated event. This tool therefore ensures that indicators of novelty and disruption used in future studies will be replicable.
 
The rationale for incorporating novelty and disruptiveness indicators in a single package comes from the fact that they both capture different aspects of the documents: the former aims at quantifying the risky profile of research, looking at the balance between exploitation and exploration \citep{march1991exploration} of the knowledge space. At the same time, the latter analyzes how impactful an article is for science. The link between novelty and citation count has been of interest in previous research \citep{uzzi2013atypical,wang2017bias} and more recently \cite{lin2021new} studied the relationship between novelty and disruptiveness indicators. The different study only looks at a specific novelty indicator, and a complete benchmark is still missing. This paper contributes to an ongoing effort to systematically benchmark and compare multiple indicators of impact and novelty by proposing an open-source tool to the community.\\

This article contributes to the Science of Science literature by providing an open source \textit{Python} package, \textit{Novelpy}, to compute Novelty and Disrutiveness measurement. It unifies in a common framework the existing indicators using a formalization based on graph theory and provides some hands-on experience. We hope that \textit{Novelpy} will contribute to homogenizing our practice in the science of science and support researchers in their work. The package will be available on \textit{Python}, one of the most popular open-source programming languages (hence with the most prominent community support) and will be maintained long term. The package currently works with a specific and documented data structure, but tools to easily use well-known data sources are under development. The package will be hosted on \textit{pypi} but also on \textit{Github}, which allows the creation of bug reporting and/or proposition of development\footnote{Documentation is available here \url{https://novelpy.readthedocs.io/en/latest/usage.html}}. The rest of the paper is structured in the following way. In Section II, we review existing programming tools for scientometrics. Section III contains the formalization of the indicators that are implemented in \textit{Novelpy}. Section IV demonstrates the package's capabilities on a random sample drawn from Pubmed. We close the paper with a discussion on the remaining limitations of novelty indicators' usages and the purpose of the package.

\section{Supported indicators}
\label{sec:indic}
    
\begin{figure}[!h]
    \centering
        \includegraphics[scale= 0.365]{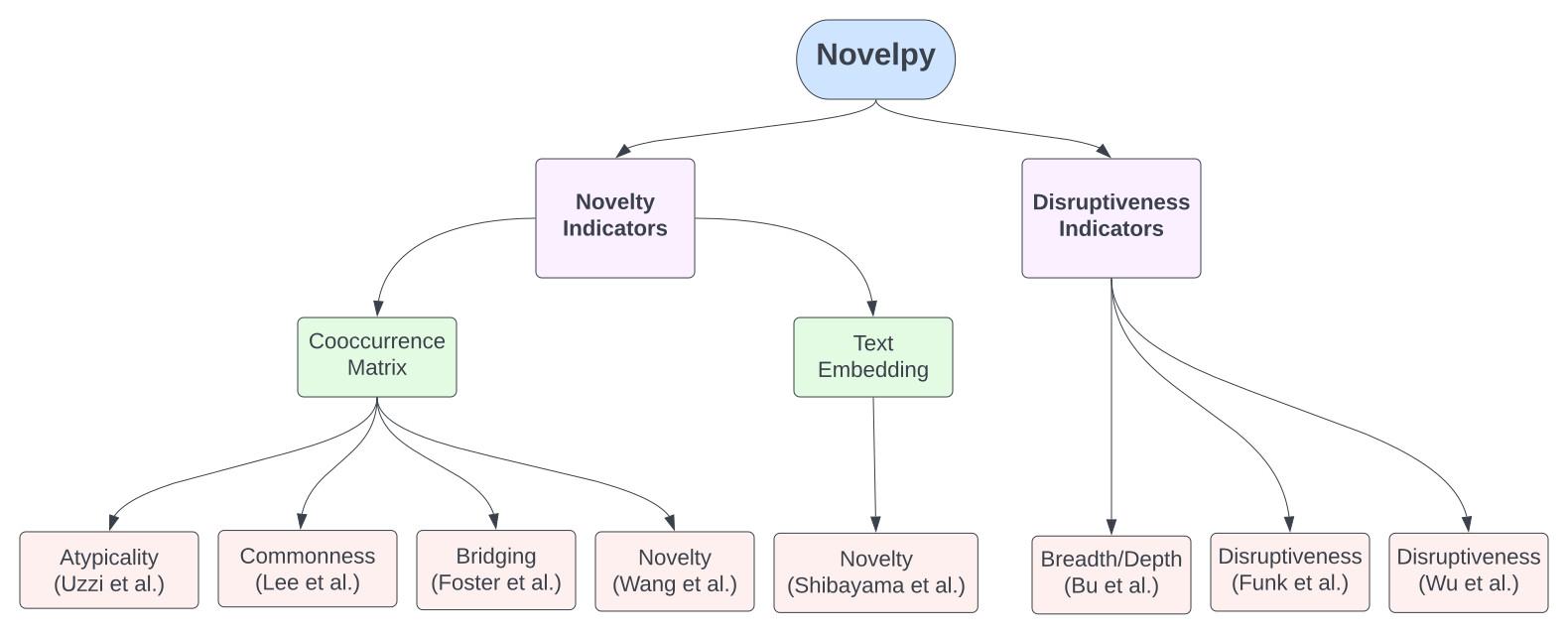}
          \caption[fig1]{\textit{Novelpy}'s module structure \footnotemark }
      \label{fig:mindmap}
    \end{figure}

    This section details the content of \textit{Novelpy}, describes the computation for each indicator, and the data required. The \textit{Novelpy} Python package provides a set of functions to perform quantitative analysis in scientometrics. The structure of the module is divided between novelty and disruption indicators. Novelty indicators are also separated between indicators based on co-occurrence matrices and ones based on text embedding techniques, as represented in figure \ref{fig:mindmap}.
    
    Practically disruptiveness indicators are all calculated through the same function, while novelty indicators have a function for each measure. All functions are explained in the module's documentation (\url{https://novelpy.readthedocs.io/}).

 Different data types can be employed depending on the indicator, as shown in \ref{tab:indicators}. All indicators working with a co-occurrence matrix can use references, journals or keywords, and disruption indices rely on the citation network. \cite{shibayama2021measuring}'s indicators use citation network and title or abstract to represent the article's semantics in a vector space. Various tools to preprocess bibliometric data are also embodied within the package to make as simple as possible computation of proposed measures\footnote{see \url{https://novelpy.readthedocs.io/en/latest/utils.html}} (e.g. co-occurrence matrix construction, text embedding, citation and authors network creation). Table \ref{tab:indicators} resume the indicators available in the module, their strengths and weaknesses, and the possible variables to compute them.

\begin{table}[!h]
\begin{adjustwidth}{-1cm}{}
    \begin{tabular}{llll|c|c|c|c|}
    \toprule
    \hline
    Type & Indicator & Pros & Cons & \multicolumn{4}{c}{Variables used}\\
    \midrule
    & & & & \rot{Ref. Journals} &\rot{Keywords} & \rot{Citation net.} & \rot{Title/Abs.} \\
    \midrule
    Novelty 
         & \cite{uzzi2013atypical} 
            & \renewcommand{\arraystretch}{0.8}\begin{tabular}{@{}l@{}}Conserve dynamical  \\ citation structure   \end{tabular} 
            & 
            \renewcommand{\arraystretch}{0.8} \begin{tabular}{@{}l@{}}Computationally\\
            intensive \end{tabular}
            & X & X & &   \\
        [15pt]
         & \cite{lee2015creativity} 
            & \begin{tabular}{@{}l@{}}Computationally lightweight \\ Data-saving \end{tabular} 
            & \renewcommand{\arraystretch}{0.8} \begin{tabular}{@{}l@{}}Conceptually\\ less advanced \end{tabular} 
            & X & X & &  \\
        [15pt]
         & \cite{foster2015tradition} 
            & \begin{tabular}{@{}l@{}}Consider undirect link \\ Computationally lightweight \end{tabular} 
            & \begin{tabular}{@{}l@{}}Discret distances  \\ \end{tabular} 
            & X & X & &  \\
        [15pt]
         & \cite{wang2017bias} 
            & \begin{tabular}{@{}l@{}}Computationally lightweight \\  \end{tabular}
            & \begin{tabular}{@{}l@{}} \\ \end{tabular}Data intensive  
            & X & X & &  \\
        [15pt]
         & \cite{shibayama2021measuring}
            & \begin{tabular}{@{}l@{}}High granularity  \\  \end{tabular}
            & \renewcommand{\arraystretch}{0.8} \begin{tabular}{@{}l@{}}Computationally and \\data intensive \end{tabular} 
            & & & O & O  \\
    \midrule
    Disruptiveness & \cite{wu2019large} & \begin{tabular}{@{}l@{}} Normalized \\  \end{tabular}  &  \begin{tabular}{@{}l@{}} Data intensive \\ Issue with term $K_{FP}$ \end{tabular} 
            & & & X & \\
        [15pt]
                & \cite{bornmann1911disruption} & \begin{tabular}{@{}l@{}} Normalized \\  \end{tabular}  & Data intensive 
            & & & X &  \\
        [15pt]
                & \cite{bu2019multi} & \begin{tabular}{@{}l@{}} Normalized \\  \end{tabular}  & Data intensive 
            & & & X &  \\
    \bottomrule
    \end{tabular}
    \caption{ \textit{Novelpy}'s available indicators. X means that you can run the indicator on either variables. O Means you need both variables to run it}
    \label{tab:indicators}
\end{adjustwidth}
\end{table}

    The module supports a wide range of data sources as long as they are in the proper format; note that transforming data to the expected structure is relatively simple. Helper functions are available to directly transform Pubmed Knowledge Graph data into the desired structure \footnote{Expected structure is presented here: \url{https://novelpy.readthedocs.io/en/latest/usage.html}
    }. For other databases, further backend to OpenAlex, Web of Science, Scopus and PATSTAT are under construction. 
    The package currently works with documents in JSON or MongoDB format. Mongo will be preferred for large databases to avoid overflowing the RAM.

\subsection{Novelty Indicators}

We focus on novelty indicators in the package based on the combinatorial idea. As discussed in the section \ref{sec:introduction}, novelty indicators can be differentiated into two groups regarding how they compute the distance between items. 
The first group uses a combination of items, such as keywords and journals, to create a co-occurrence matrix. Algorithms make use of this matrix to compute the distance. The more distant, the more unexpected and, therefore, novel is the combination. The second type of indicators maps items in a Euclidian space with text embedding techniques like word2vec \citep{mikolov2013distributed}. The distance is then computed in this semantic space. 
As shown in Figure \ref{fig:mindmap} novelty indicators are split between those using co-occurrence of entities such as journals or keywords and those using word embedding techniques. For the first group of indicators, we first need to create a co-occurrence matrix for each year of the given dataset. While some indicators only use the focal year to compute the score for each combination \citep{uzzi2013atypical, lee2015creativity, carayol2019},  others take into account past combinations in the score calculation \citep{foster2015tradition} and future re-utilisation \citep{wang2017bias}.

Atypicality \citep{uzzi2013atypical}, Commonness \citep{lee2015creativity}, and Novelty \citep{wang2017bias} are all indicators that use references of an article at a journal level. Previous studies usually focused on one type of knowledge unit, but as long as one can create a co-occurrence matrix between items, it becomes trivial to generalize. \cite{carayol2019} reformulate \cite{lee2015creativity} and apply it to keywords and construct the indicator accounting for inter-field heterogeneity by splitting the analysis. \cite{fleming2001} computes a combination of patent subclasses, a prevalent practice in patentometrics.
\cite{dahlin2005invention} propose a novelty measure based on the overlapping between documents references that was reused by \cite{trapido2015novelty}, based on this work \cite{matsumoto2021introducing} propose an extension that computes the average share of references that are shared between a focal paper and all other documents in the same field. These indicators are not present in \textit{Novelpy} (v1.2) but will be added in future versions.

Although the co-occurrence matrix can be considered an adjacency matrix, only a handful of indicators use graph theory to compute the distance between items. Indeed indicators \textit{à la} \cite{uzzi2013atypical}, or \cite{lee2015creativity} take into account only the direct neighbourhood during distance calculation. If items A and B are close, items B and C are close, and D is not related to any of them, then the combination of A and C is more likely to happen than A and D. This logic is completely ignored if one considers the direct neighbours. \cite{wang2017bias} integrated this into their indicator by considering the cosine similarity between nodes' neighbours, it takes into account common friends (A and C in the example above). Using community detection as in \cite{foster2015tradition}, one can better represent the distance between two units by using the global structure of the network. However, the discrete nature of the novelty score can be argued. 
Using text embedding, one can have a continuous representation of the distance between items. This distance is related to the text's structure since word similarity depends on their neighbourhood. Some initiatives used these techniques with different purposes but could be used to create a novelty score.
 \cite{hain2020text} create a similarity measure between patent using word2vec \citep{mikolov2013distributed}. \cite{shibayama2021measuring} was the first to apply word embedding techniques in a novelty context. They embed references in a Euclidian space using spaCy and then compute a distribution of cosine distances between documents present in the references for a given document.

\vspace{0.5cm}

We propose a mathematical formalization of these indicators. By setting up this framework, we offer a basis for defining future new indicators. These indicators are formulated based on graph theory, where the network's nodes are units of knowledge (journals, keywords or references) and edges represent the co-occurrence of these units in entities (documents or patents).

\begin{itemize}
\item Co-occurrence matrix can be written as a graph $G=(V,E,w)$. 
\item Set of nodes $V$ of dimension $v$ represent here the entities (e.g. keywords, journals), a given entity is defined as $V_i$.
\item Set of edges is noted $E$.
\item Number of combinations between $V_i$ and $V_j$ is the weight for the edge $(V_i,V_j)$ and is written $w({V_i,V_j})$.
\item Degree of a node $V_i$ is written $k_i$. $N$ is the sum of the weighted edges in $G$ without self-loops, $N=\Sigma_{i=1}^{v-1}\Sigma_{j=i+1}^{v}w(V_i,V_j)$.

\end{itemize}

$D$ define our set of documents of dimension $n$. Each focal paper, $FP$, has its network, which can be defined as $G_{FP}$, $E_{FP}$ is the subset of edges present in document $FP$. $G_{FP}$ uses the same set of nodes $V$ as $G$ and can be express as $G_{FP}=(V,E_{FP},w_{FP})$. In some cases, $G_{FP}$ is an unweighted network and will be written then $G_{FP}=(V, E_{FP})$. The number of links, $w(V_{i}, V_{j})$, is then defined as the sum of all combinations of two given entities overall document in $D$, $w({V_i, V_j})=\Sigma_{d=1}^{n}w_{d}(V_{i}, V_{j})$ where $w_{d}(V_{i}, V_{j})$ is binary if the graph is unweighted at the document level. 
$G(V,E,w)$ can be split at a year level. For example, in year $t$, and the associated network will be noted $G_t(V, E_t,w_t)$. \cite{uzzi2013atypical}, \cite{lee2015creativity}, use only the subgraph $G_t$ for calculation. \cite{foster2015tradition} use the accumulation of past networks. For \cite{wang2017bias}, several subgraphs are involved in computing the indicator. The novelty indicators \textit{à la} \cite{wang2017bias} deal with four subgraphs of $G$. One needs to consider two different past sets of documents (noted $P$ and $B$) and a set of future documents (noted $F$).

\subsubsection{\cite{uzzi2013atypical}: Atypicality}

\begin{figure}
\centering
\begin{subfigure}{.5\textwidth}
  \centering
  \includegraphics[scale= 0.4]{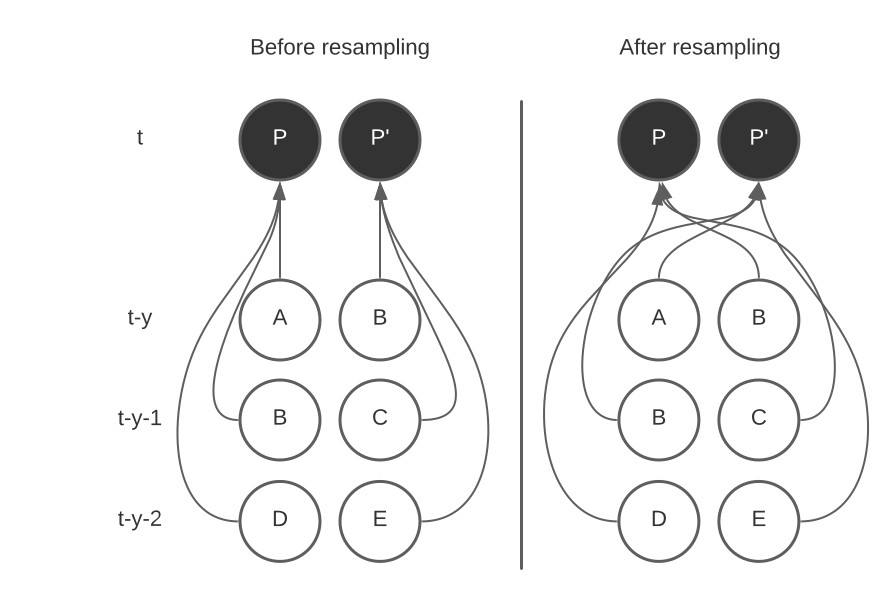}
  \caption[Fig2] {Resampling strategy}
\end{subfigure}%
\begin{subfigure}{.5\textwidth}
  \centering
  \includegraphics[scale= 0.4]{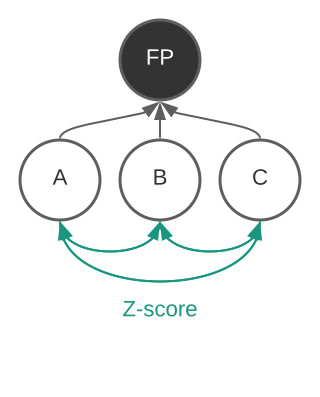}
  \caption{Score calculation}
\end{subfigure}
\caption[Fig1] {\cite{uzzi2013atypical} \footnotemark}
\label{fig:2}
\end{figure}

The goal of the measure proposed by \cite{uzzi2013atypical}, called ``Atypicality'', is to compare an observed network with a random network. The network is shuffled, preserving the temporal distribution of references at the paper level. As shown in Figure \ref{fig:2} a document citing two articles from, say, 1985 and one from 1987 will still cite articles published the same year, but the journal can change. The frequency of the combination $(V_i, V_j)$ at time $t$ is defined as  $w_t(V_i, V_j)$, and we extract the adjacency matrix of observed frequencies. The idea is basically to compute the frequency Z-score for each journal combination. The Z-score is defined as $z=(obs-exp)/\sigma$; an observed frequency is compared with a theoretical one. 

The theoretical frequency is generated through Markov chain Monte Carlo simulation, preserving the dynamical structure of citations. In the case of Atypicality, we are dealing with $s+1$ different networks for the year $t$, the existing network and $s$ resampled ones. The existing network is $G_t$, as defined above. The others are generated by preserving the temporal distribution of references in an article. For each document $FP$ we want to keep the number of references published in year $t-y$
stable for all $y$ to ensure that the global age distribution of the pieces of knowledge used at time $t$ remains stable.

One needs to generate $s$ random networks $G_{t}$. After re-sampling, the publishing year of references is no anymore taken into account. Edges' weights are then aggregated to fit with $G_t$ edge structure $E_t$ by summing over all combinations. The observed frequency for each sample is computed for each edge $(V_i, V_j)$. We write the set of frequencies for the combination of $V_i$ and $V_j$ in the $s$ samples $w_t^s({V_i, V_j})$. One can then compute the mean and standard deviation for each edge's frequency and compute a z-score.

\footnotetext{(a): P and P' are two distinct papers, P cites journals A, B and D. P' cites journals B, C and E. The goal is to shuffle the network by conserving the dynamic structure of citations at the paper level. P is no longer citing A from $t-y$ but cites B from year $t-y$. (b): Comparing the observed and resampled networks, we can compute a z-score for each journal combination. }

$$Z-score_{ijt}=\frac{w_t({V_i,V_j})-mean(w_t^s({V_i,V_j})}{std(w_t^s({V_i,V_j}))}$$

For each paper, taking all combinations made ($E_{FP}$), a distribution of z-score written $Z_{FP}$ is computed, and the 10th percentile ($P_{10}$) of this distribution (the novelty) and the median ($P_{50}$) (the conventionality). The novelty and conventionality for document $FP$ are then written:

$$Novelty_{FP} = P_{10}(Z_{FP})$$
$$Conventionality_{FP} = P_{50}(Z_{FP})$$

While this indicator only requires data from a specific year, it is still computationally greedy. Indeed generating the $s$ samples and the computation of the average and the standard deviation for each possible combination is expensive. On the contrary, this indicator allows for keeping the temporal structure stable, which is more in line with the reality of the availability of the knowledge pieces.

\subsubsection{\cite{lee2015creativity}: Commonness}

\cite{lee2015creativity} compare an observed network with a theoretical network (Observed vs Expected frequency of edges) at a year level. The observed number of combination $(V_i, V_j)$ at time $y_t$ is the number of edges w $w_t(V_i, V_j)$, the theoretical number of combination is $\frac{k_i*k_j}{N_t}$, the degree for entity $i$ and $j$ multiply together and divided by the total number of combination made in year $t$.

\begin{figure}[h!]
  \centering
  \includegraphics[scale= 0.4]{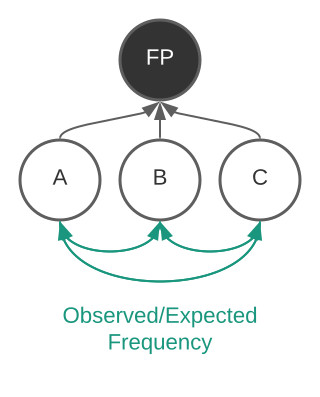}
  \caption{\cite{lee2015creativity}}
\end{figure}

$$ Commonness_{ijt} = \frac{w_t(V_i,V_j)*N_t}{k_i*k_j} $$

For each paper, taking all combinations made in document $FP$ ($E_{FP}$), a distribution of commonness-score written $C_{FP}$ is computed. The commonness for document $FP$ is the 10th percentile ($P_{10}$) of this distribution and is written as:

$$Commonness_{FP} = -log(P_{10}(C_{FP}))$$

The main advantage of the commonness indicator is its speed of calculation; it is the minor demanding indicator in terms of the execution time of the package. The indicators only require data from a specific year. Note that this indicator is very close to \cite{uzzi2013atypical}'s one. Both would be equal if \cite{uzzi2013atypical} resampling method would not consider the references' publishing year.

\subsubsection{\cite{foster2015tradition}: Bridging}

\cite{foster2015tradition} propose a novelty indicator based on community detection algorithms. It captures the distance between two entities taking into account undirected edges. The goal of the measure is to identify the network's community studied and capture proximity through the community in which the combined entities are clustered. 

Any community algorithm can be applied to this indicator. We rely on the Louvain algorithm in \textit{Novelpy} following\cite{foster2021surprise}, but we intend to add further options. After applying the community algorithm on $G(V, E,w)$, we are left with multiple clusters of entities. $C_i$ is the community to which the entity $i$ belongs.

\begin{figure}[h!]
  \centering
  \includegraphics[scale= 0.4]{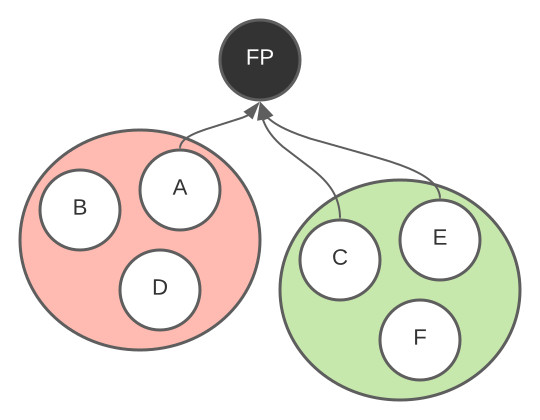}
  \caption[Fig3]{\cite{foster2015tradition} \footnotemark}
\end{figure}
\footnotetext{FP cites different journals which belong to different communities. The novelty is the number of journals' combinations from two different communities. Communities of journals are computed through community detection algorithm.}

$$Novelty_{FP} = \frac{\sum\limits_{(i, j) \in E_{FP}} 1-\delta(C_i,C_j)}{|E_{FP}|} $$

Where $\delta(C_i, C_j) = 1$  if $C_i = C_j$ (i.e. both entities, i and j, are in the same community), $\delta(C_i, C_j) = 0$ otherwise. The novelty score of an entity is the proportion of pairwise combinations that are not in the same community.

This indicator brings into the field algorithms that capture global network structure and only require data from a specific year. At the same time, this indicator does not allow measuring distances between communities and proposes only a binary distinction.

\subsubsection{\cite{wang2017bias}: Novelty}

\cite{wang2017bias} propose a measure of difficulty on pair of references never made before. These new pairs need to be reused after the given publication's year (Scholars do not have to cite directly the paper that creates the combination but only the combination itself). The idea is to compute the cosine similarity for each journal combination based on their co-citation profile $b$ years before $t$. The cosine similarity between $W^B_i$ and $W^B_j$ is defined : 

$$COS(W^B_i,W^B_j) = \frac{W^B_i.W^B_j}{\|W^B_i\| \|W^B_j\|}$$
where $W^B_i$ represent all links of entity $i$, $B$ years before year $t$.

Novelty \textit{à la} \cite{wang2017bias} relies on four subgraphs of $G$ constructed using two different past sets of documents, a set of future documents, and the set of documents for the focal year. These different subgraphs are defined as follows (Note the first year of the data set $y_0$ and the last as $y_n$) :
\begin{itemize}
    \item $G_t=(V,E_t,w_t)$ is a subgraph of $G$ from year $t$ (documents published year $t$)
    \item $G_P=(V,E_P,w_P)$ is a subgraph of $G$ from year $t_0$ to $t-1$ (documents published before year $t$)
    \item $G_{B}=(V,E_{B},w_{B})$ is a subgraph of $G$ from year $t-b$ to $t-1$ is used to measure the cosine similarity between nodes. This set is a subgraph of $G_P$ (documents are published in a given window before year $t$)
    \item $G_F=(V,E_F,w_F)$ is a subgraph of $G$ from year $t+1$ to $t+f$ (documents published in a given window after year $t$)
\end{itemize}

\begin{figure}[h!]
  \centering
  \includegraphics[scale= 0.4]{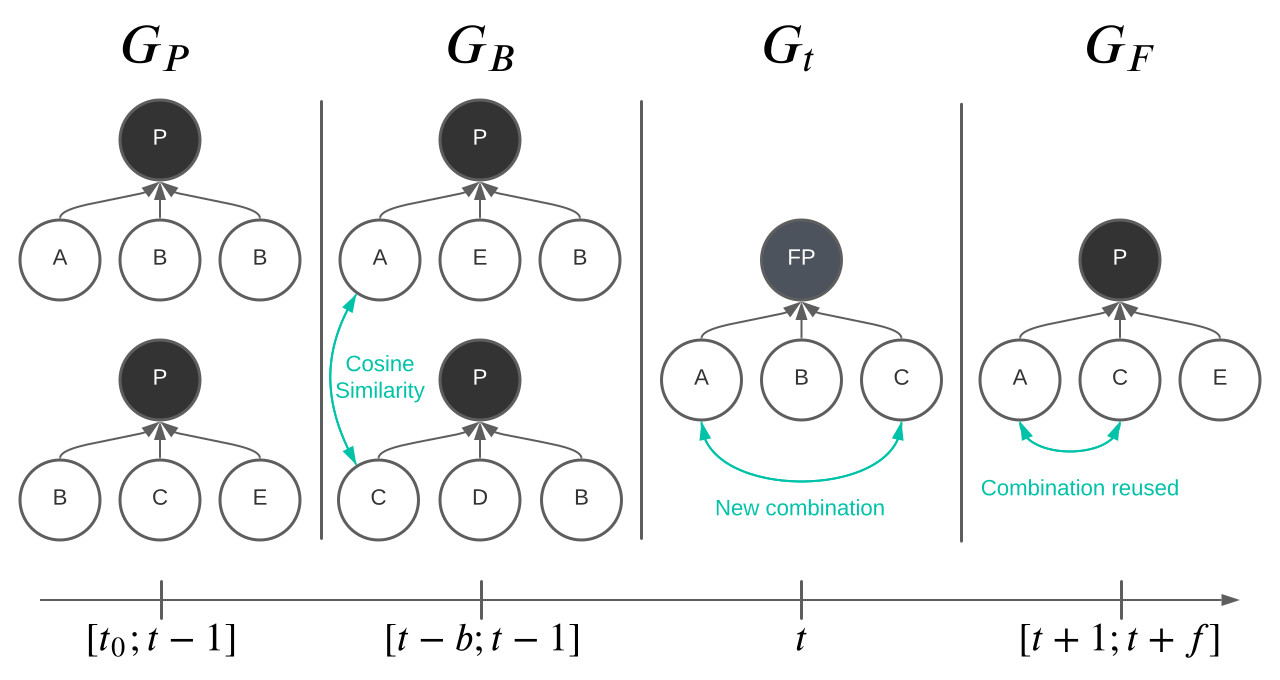}
  \caption[fig4]{\cite{wang2017bias} \footnotemark}
\end{figure}
\footnotetext{For a given article at time $t$, we check if the journal combined were already combined in the past ($G_P$). We then check if the combination is reused in the future ($G_F$). If the combination is new and reused, the difficulty of making such a combination is calculated on the recent past ($G_B$)}

This indicator focuses on new combinations reused afterwards and not achieved before the given year $y_t$. One needs to keep all elements of $E_t \notin E_P$ and $E_t \in E_F$. More precisely, edges belonging to the following subset (that we call $E_N$) are the only edges used to compute this indicator $E_N = (E_t \cap E_F) \cap \overline{E_P}$

Cosine similarities are calculated using $G_{B}$. For each document, we compute an undirected and unweighted network. The novelty is the sum of all edges from $E_{FP} \in E_N$, that is:

$$ Novelty_{FP} = \sum\limits_{(i, j )\in E_N}1-COS(W^B_i,W^B_j)$$

The main issue with this indicator is the amount of data needed to compute the measure. One needs as much data as possible before the focal year to ensure that the combination has never been made. At the same time, some hyperparameters involved in this measurement can drastically modify the results. For example, the time window to capture the re-utilisation of a combination or the number of times reused needed to be novel are very arbitrary.

\subsubsection{\cite{shibayama2021measuring}: Novelty}

\cite{shibayama2021measuring} propose to incorporate semantic distances to capture diversity in the set of references from a given article following \cite{hain2020text} and their similarity measure between patents. Documents centroids are computed by summing all words representation for each document.

Consider a directed unweighted graph $G(V, E)$ containing the citation network. For a given document $FP$, a referenced document is noted $r$, and the set of nodes that are cited by $FP$ is then $In_{FP} = \{r : (FP,r) \in E \} $. \cite{shibayama2021measuring} compute all distances between each documents centroids ($C_{|In_{FP}|}^{2}$ combinations). All documents have a vectorial representation in a semantic space of length 200. Distances between two references $i, j \in In_{FP}$ are calculated through cosine similarity: $n_{ij} = 1-COS(T_i, T_j)$ where $T_i$ is the dense vector text representation for a document $i$. A distribution of novelty score $N_{FP} = \{n_{ij}: i,j \in Out_{FP}\}$ is then computed, for each document, the final score is a percentile of $N_{FP}$

\begin{figure}[h!]
  \centering
  \includegraphics[scale= 0.4]{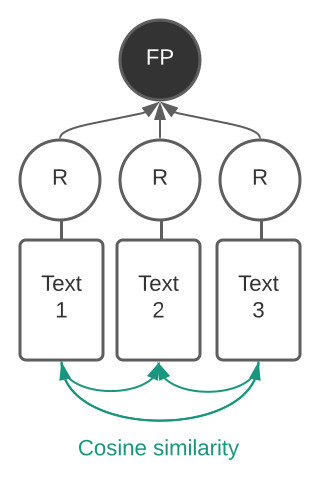}
  \caption[Fig5]{\cite{shibayama2021measuring} \footnotemark}
\end{figure}
\footnotetext{For a given article, each reference's abstract (or title) is represented in a semantic space through text embedding techniques. The distance between two references is then computed through cosine similarity. }
$$Novelty_{FP} = P_{q}(N_{FP})$$

\cite{shibayama2021measuring} indicators is both data-intensive and computationally intensive. One needs to get all references' titles/abstracts for a given set of articles.
The package currently works with a word2vec pre-train \textit{en\_core\_sci\_lg} from \textit{spacy} to compute the dense representation of a document. Future versions will incorporate a back-end to use any pre-train.

\subsection{Disruptiveness Indicators}

Disruptiveness indicator offers alternative measures of impact to the number of citations. They allow understanding if a given article behaves as a bottleneck between the knowledge mobilized in a given article and the articles that will cite it. It has been introduced in scientometrics by \cite{wu2019large} and was previously proposed
for patents by \cite{funk2017dynamic}. Following \cite{azoulay2019small} definition, a paper can either consolidate existing knowledge or disrupt it. Future papers that cite a focal paper and its references do not use fundamental new pieces of knowledge created in it (i.e. the focal paper consolidates the existing knowledge space but does not disrupt the playing field). On the other hand, if future papers cite only the focal paper and not its references, then the focal paper is considered disruptive. Quoting \cite{bornmann1911disruption} ``[...] many citing documents not referring to the FP’s cited references indicate disruptiveness. In this case, the FP is the basis for new work which does not depend on the context of the FP, i.e. the FP gives rise to new research.''
All presented measures normalize citation and give relative perspective to publication's impact  \citep{bu2019multi}. Disruptiveness indicators consider the importance of pieces of knowledge (references) in a given article for other articles, whereas Depth and Breadth, as proposed in \cite{bu2019multi}, capture how the knowledge generated by that given item is reused and whether it allows for the consolidation of a domain or is instead used in a disparate manner. 

Consider a directed unweighted graph $G(V, E)$ containing the citation network.
\begin{itemize}
    \item For a given document $FP$ we note a document cited by $FP$, $r$. The set of nodes that are cited by $FP$ is then $In_{FP} = \{r \in V | (FP,r) \in E \} $
    \item For a given document $FP$ we note a document citing $FP$, $c$. The set of nodes that are citing $FP$ is then $Out_{FP} = \{c \in V\ | (c,FP) \in E \} $
    \item The number of citations for $FP$ is then $deg^-(FP) = |Out_{FP}|$ and number of references $deg^+(FP) = |In_{FP}|$
    \item The set of references for an article citing $FP$ is then noted $In_c$
\end{itemize}

\subsubsection{\cite{wu2019large}: Disruptiveness}
\label{sec:wu}
By adapting \cite{wu2019large} notation, we called $I_{FP}$ the set of nodes with $FP$ as a parent that does not have $FP$'s parents as parents. More formally $I_{FP} = \{c \in Out_{FP} | In_c \notin In_{FP}\}$. The set of $J_{FP}^l$ is the set of nodes with $FP$ as a parent that share at least $l$ parents with $FP$. We note $J_{FP}^l = \{c \in Out_{FP} | |\{In_c \in In_{FP}\}| > l \}$. Finally, $K_{FP}$ is the set of nodes that share parents with $FP$ but that do not have $FP$ as a parent: $K_{FP} = \{v \in V | v \in In_{FP}\}$.

\begin{figure}[h!]
  \centering
  \includegraphics[scale= 0.4]{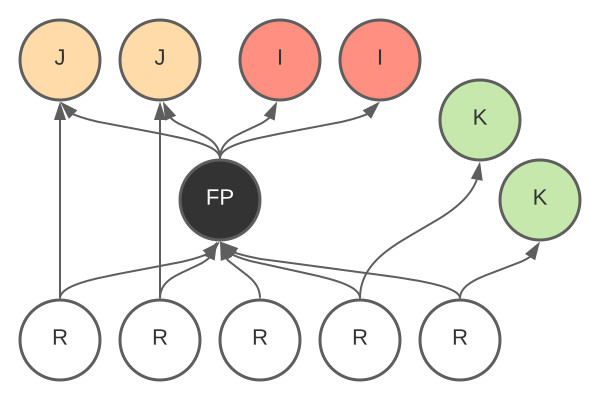}
  \caption[Fig8]{\cite{wu2019large,bornmann1911disruption} \footnotemark}
\end{figure}
\footnotetext{For a given article FP, we retrieve:  (a): Articles citing FP and references from FP (named J). (b): Articles citing FP but no references from FP (named I). (c): Articles citing references from FP but do not cite FP (named K). }

The disruptiveness \textit{à la} \cite{wu2019large} is then noted :
$$DI_1 = \frac{|I_{FP}|-|J_{FP}^1|}{|I_{FP}|+|J_{FP}^1|+|K_{FP}|}$$

Some variants that consider only paper sharing at least $l$ references have been proposed:
$$DI_5 = \frac{|I_{FP}|-|J_{FP}^5|}{|I_{FP}|+|J_{FP}^5|+|K_{FP}|}$$

\subsubsection{\cite{bornmann1911disruption}: Disruptiveness}

A variant that removed the term $|K_{FP}|$ has been proposed by \cite{wu2019solo} because the number of documents that cite references from the focal documents without citing the focal documents is often too large compared to the paper from other sets. \cite{wu_wu_2019} show how considering the set $K_{FP}$ can lead to decrease the disruptiveness when the term $|I_{FP}|-|J_{FP}^1|$ is negative. In that configuration, more papers that do not cite $FP$ ($|K_{FP}|$) lead to a higher disruptiveness, which is different from how the indicators conceptually work.
Defined as $DI_l^{no k}$ by \cite{bornmann1911disruption}, we note:
$$ DI_l^{no k} =  \frac{|I_{FP}|-|J_{FP}^l|}{|I_{FP}|+|J_{FP}^l|} $$

\subsubsection{\cite{bu2019multi}: Breadth and Depth}

\cite{bu2019multi} propose an alternative to the disruptiveness indicators described above. It calculates the proportion of articles citing the focal paper that also cites other articles citing it. The indicator allows us to understand whether the document contributes to a restricted search domain; the documents citing the focal paper are interdependent and cite each other. On the contrary, the documents using the focal paper research may also be unconnected and belong to more extensive research space. 

\begin{figure}[h!]
  \centering
  \includegraphics[scale= 0.4]{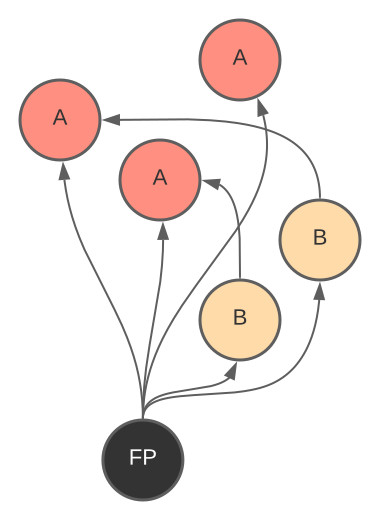}
  \caption[Fig9]{\cite{bu2019multi} \footnotemark}
\end{figure}
\footnotetext{For all articles citing FP, we check if they also cite papers citing FP. }

Let $FP$ be the focal paper, the articles citing it the set $Out_{FP}$. We are interested in the articles cited by the documents of the set $Out_{FP}$. For each element $c$ of $Out_{FP}$, we observe a set of associated references named $In_c$.
The proportion of documents citing document $FP$ and also citing documents that are citing $FP$ is then written as: 

$$Depth_{FP} = \frac{|\{c \in Out_{FP} : |In_c \in Out_{FP}| >0\}|}{|Out_{FP}|}$$

On the contrary, the breadth, the proportion of papers citing $FP$ that do not cite other publications also citing $FP$, is written:

$$Breadth_{FP} = \frac{|\{c \in Out_{FP} : |In_c \in Out_{FP}| =0\}|}{|Out_{FP}|} = 1 - Depth_{FP}$$

\cite{bu2019multi} also propose a dependence measure. It captures the average number of references shared between the focal paper $FP$ and documents citing it. $In_{FP}$ is the set of references of $FP$. For all document $c$ that cite $FP$ ($Out_{FP}$), we want to know the number of references shared:$|\{ In_c \in In_{FP}: c \in Out_{FP} \}|$. The average number of references shared between document $FP$ and all documents citing it ($c \in Out_{FP}$) is then:

$$Dependence_{FP} = \frac{\sum\limits_{c \in Out_{FP}}| In_c \in In_{FP}|}{|Out_{FP}|}$$

Two other indicators from \cite{bu2019multi} are not computed in our function: Independence and Dependence. However, they represent the proportion of publications citing a focal paper that also cites references from the focal paper. Using notation from \ref{sec:wu}: $\frac{|I_{FP}|}{|I_{FP}|+|J_{FP}^1|}$ one can easily derive this value from disruptiveness indicators $DI_1^{no k}$. Indeed from $DI_1^{no k} =  \frac{|I_{FP}|-|J_{FP}^1|}{|I_{FP}|+|J_{FP}^1|}$ we can compute the independence, the proportion of articles citing the focal paper that do not cite articles cited by the focal paper $$\frac{|I_{FP}|}{|I_{FP}|+|J_{FP}^1|} = \frac{DI_1^{no k}+1}{2} \quad\textrm{if}\quad |Out_{FP}| > 0$$

All these measures are pretty heavy in terms of data requirements. Indeed, for each given article, we need to access the references, the articles citing the focal paper and the articles citing the references of the focal paper.

\section{Sample analysis}
\subsection{Descriptive statistics}
\begin{figure}[H]
\centering
\begin{subfigure}{.4\textwidth}
  \centering
  \includegraphics[width=1\linewidth]{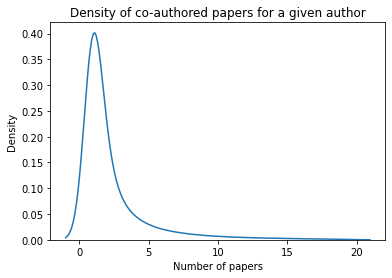}
  \caption{}
  \label{fig:sub1}
\end{subfigure}%
\begin{subfigure}{.4\textwidth}
  \centering
  \includegraphics[width=1\linewidth]{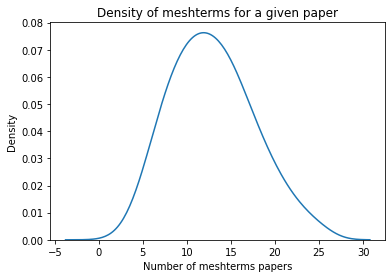}
  \caption{}
  \label{fig:sub2}
\end{subfigure}
\begin{subfigure}{.4\textwidth}
  \centering
  \includegraphics[width=1\linewidth]{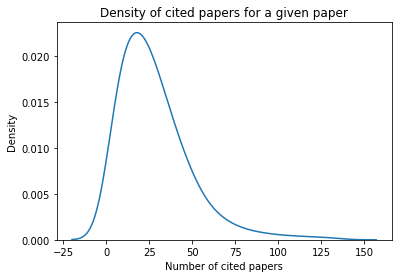}
  \caption{}
  \label{fig:sub3}
\end{subfigure}
\caption{(a) Density of contribution of authors. On average an author has 2.6 publications (solo or co-authored) in 10 years. (b) Density of the number of mesh terms between 2000-2010. On average, a paper is labelled with 13 mesh terms. (c) Density of references between 2000-2010. On average, a paper has 23 references }
\label{fig:10}
\end{figure}

This section provides examples of applications which could be done with \textit{Novelpy}. We use Pubmed Knowledge Graph (PKG) sample \citep{xu2020building}, which stores the research articles published on Pubmed and offers metadata for all papers. This analysis is proposed as an example to show our module features after computing the indicators. Interested readers will find code and resources to create tables, plots and indicators here \url{https://novelpy.readthedocs.io/en/latest/usage.html\string#id5}. Every figure and table can be found in Annex. The sample is restricted from 1995 to 2015, and the focal period is 2000-2010. The sample is composed of 1,469,352 papers and 2,959,650 distinct authors. Authors are disambiguated in PKG using advanced heuristics and algorithms. The sample was done so that every article has the attribute we need to run the indicators. Each paper has a list of references, mesh terms, authors, title and abstract. Table \ref{tab:table2} and Figure \ref{fig:10} summarize the statistics of the sample. On average, the number of references used in a paper is 23, which is coherent with typical citation behaviour \citep{abt2002relationship}.
On the other hand, having a sample of a Database restricted to health science reduces the number of papers that an author contributed to during ten years (\ref{fig:sub1}). This issue highlights a potential bias when creating indicators based on references and authors. The number of papers almost doubled in 10 years, which is coherent with the literature \citep{fortunato2018science}.

\subsection{Results}

\begin{figure}[H]
  \centering
  \includegraphics[scale=0.25]{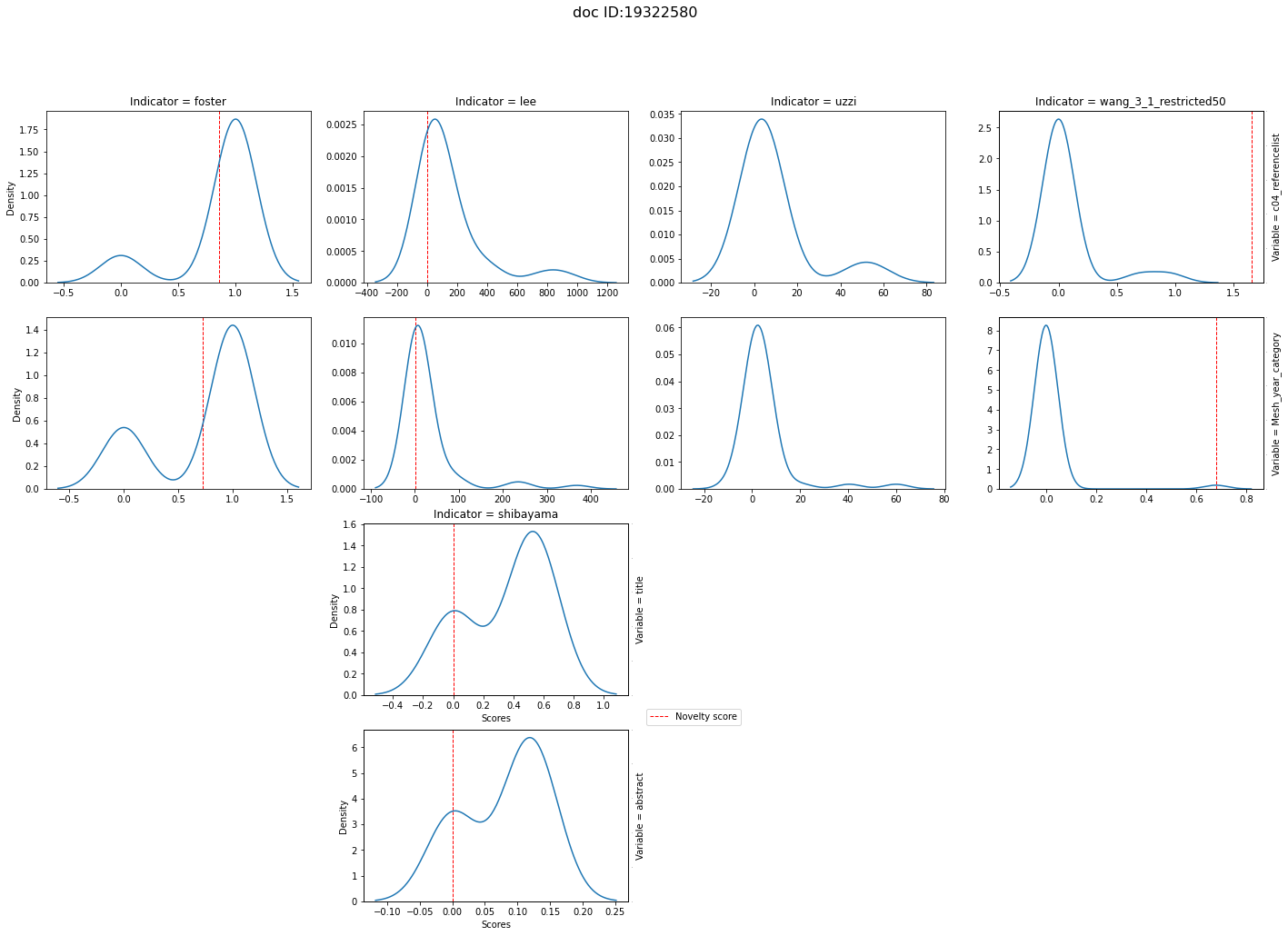}
\caption{Each combination has a novelty score. A single plot represents the density of score combinations for a specific paper (PMID 10698680) for a specific indicator and entity (i.e. mesh terms, journals, title, abstract). The scores for the first row were computed using a combination of cited journals. The scores for the second row were computed on mesh terms combinations. Each column represents an indicator. The last two rows are for text embedding-based indicators (i.e. \cite{shibayama2021measuring}: Novelty, Author proximity) on the title of the paper or abstract.}
\label{fig:11}
\end{figure}

\begin{figure}[H]
  \centering
  \includegraphics[scale=0.25]{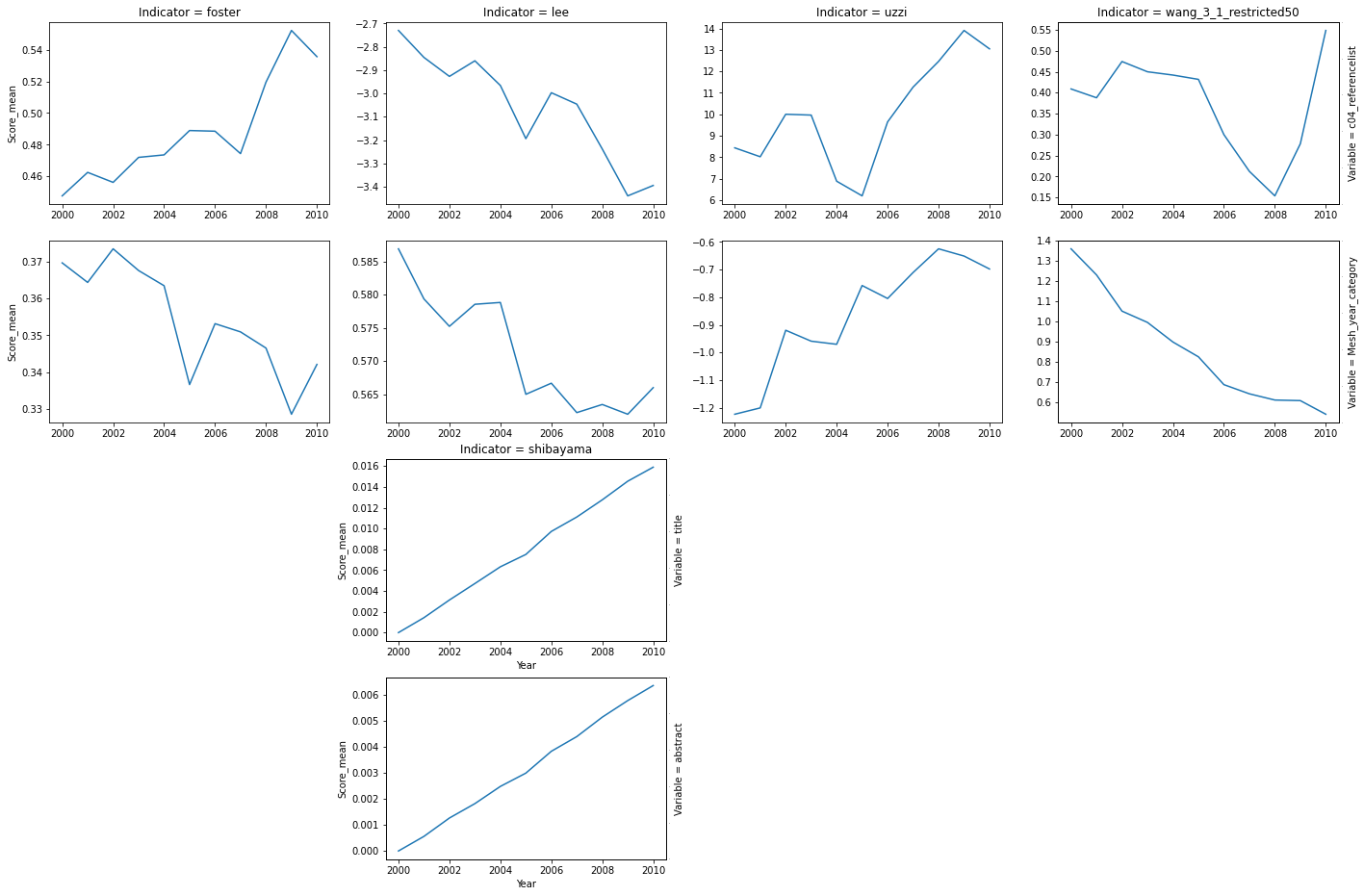}
\caption{The mean novelty score on every document for a given year. Columns and rows represent respectively indicators and variables.}
\label{fig:12}
\end{figure}

\begin{figure}[H]
  \centering
  \includegraphics[width=0.50\linewidth]{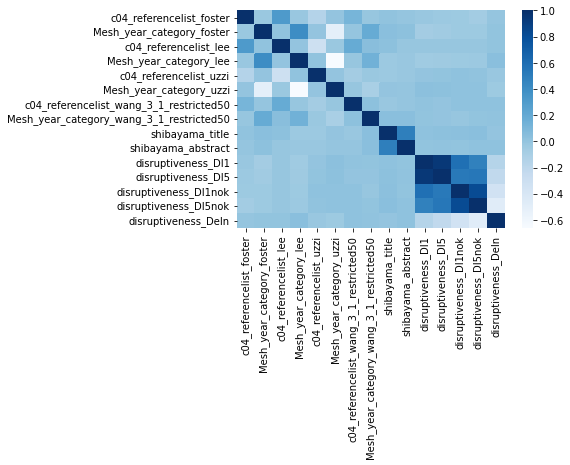}
\caption{The correlation between the novelty score for each indicator, given the entity, for the period 2000-2010}
\label{fig:13}
\end{figure}

As discussed in previous sections, research on novelty indicators still needs to be conducted on multiple axes. \textit{Novelpy} will help compute different indicators on different entities. Researchers can then use the novelty scores given by the package to perform their analysis. Individual level analysis can be done by looking at the distribution of novelty score as in Figure \ref{fig:11}. Comparing indicators and studying the evolution of novelty over the years are the main motivations of this package.Only a handful of studies look at the dynamics of novelty over time.
Nevertheless, understanding the evolution of creativity in papers, patents, or other entities can provide insight into the trade-off between exploration and exploitation of the research space in a given field. Figure \ref{fig:12} shows the evolution of the novelty score mean for each indicator given the variable (references, mesh terms). We can not draw conclusions since the sample is random and aggregated over all the fields inside Pubmed. The pattern of trends is highly different depending on the indicator and variable. This heterogeneity could be the evidence that further investigation is needed to understand what precisely these indicators capture and in which case they predict novelty the best. This question is even more relevant considering the lack of correlation of indicators in Figure \ref{fig:13}. 

\section{Discussion}

This paper is meant to demonstrate the capabilities of the new \textit{Python} package \textit{Novelpy}. We proposed a sample analysis using the functions in this package to put forward how it can help interested readers either compute and analyze already existing indicators or tackle current problems related to novelty measurement. Several criticisms can be made on current novelty measurement and are crucial points to investigate to solidify our knowledge and usage of these indicators.

The diversity and the convergence in how novelty indicators are created raise inquiries on what they measure. As seen in our sample analysis, the results are highly dependent on the indicator used, which confirms the previous interrogation on cherry-picking the indicator \citep{shibayama2021measuring,foster2021surprise}. At the same time, indicators are often used on the same entity (keywords or references' journal). Recent measures like \cite{shibayama2021measuring} and \cite{arts2021natural} expand this domain by using text information of references. Novelty indicators are rarely conceptualized and often need a qualitative background. Qualitative study like \cite{tahamtan2018creativity} questions the importance of literature in authors' creative process. The link between references and creativity is debated; one should investigate if references can be used safely as a proxy variable for creativity. 

Research evaluation was done only by experts in the Scientometric field and specialists working for public institutions. Open access data created a recent trend to entrust this evaluation to more diverse researchers and public workers. New actors need to have the necessary tools to compute scientometric indicators and at least some knowledge of the relevance of said indicators. Using software creates a gap between the user and the actual data. This deficiency may raise some issues if assumptions necessary for the relevance of the indicators are omitted. A data-driven decision can become inefficient because the algorithm used is a black box and was misused. A solid background on how and why these indicators are created is necessary to limit bias in selecting indicators when used in research. As seen in section 3, every indicator has its pros and cons and different hyperparameters (time window, re-utilization, number of samples, and others) and highly depends on the database used. The coverage is highly heterogeneous depending on the database used (language, fields, nationality, and others) \citep{sugimoto2018measuring}. These aspects and the increasing number of novelty indicators create an arbitrary decision when using them. \cite{sugimoto2018measuring} suggests that indexing and classification of documents differ between databases, making it challenging to reproduce studies on other databases. It is difficult to construct a general indicator that can be applied to all scientific disciplines, as citation habits are heterogeneous, making comparisons between fields risky \citep{carayol2019} proposed to compute scores by field, but this is not the norm). Depending on the country, methods and standards may differ within a discipline, and the historical practice of a field may change the representations.

Improving novelty measurement is essential to support innovative research. Highly novel documents are less likely to be cited in the short run and are less likely to be published in journals with high impact factor \citep{wang2017bias, mairesse2021impact}. Because of this pressure from citation count evaluation, exploration of the science is less likely to be done. Researchers will tend to conform to conventional references of her field, which is already accentuated during the submission process. Documents that are already highly cited, considered like a stepping stone in the field, will therefore get even more cited, creating a vicious circle. This vicious circle has the consequence of narrowing the research where only those that agree with the already existing paradigm will be rewarded with citation. This phenomenon is already seen in AI research where topics get less and less diverse \citep{klinger2020narrowing}. The goal of science is not to persist in a good-enough solution but to explore different possibilities, even fruitless ones. Citation indicators usually do not highlight researchers taking risks by trying something new. Different funding methods to bolster High-Risk/High-Reward (i.e. highly novel) research already exist \citep{ocde}. Experts are not free of bias when evaluating novelty, funding processes are not homogeneous, and many decisions remain arbitrary. None of them uses novelty indicators to evaluate the proposal. Novelty measurement may be relevant to offer reliable information when handing over grants to a research proposal.

We finish this discussion with a roadmap and our hopes for \textit{Novelpy}. The main missing feature we want to develop in future versions is the automatic run using well-known Databases (PATSTAT , Microsoft Academic Knowledge Graph, Arxiv,...). Currently, users need to pre-process data to match our format. Although we give a complete example and made the sample available here \url{https://novelpy.readthedocs.io/en/latest/usage.html\string#id5}, we do believe that expanding the accepted inputs will help researchers work on improving novelty indicators. The second feature we want to add is a time complexity analysis. In order to do a proper benchmark between indicators, we need to compare the computing speed. Users can currently do it manually, but we intend to facilitate the process and add plots to fill this gap. Finally, we will add, sparingly, new and past indicator. Anyone interested in working on the module can head to GitHub \url{https://github.com/Kwirtz/novelpy} and create a pull request.

\bibliography{ref_novelpy}
 
 \pagebreak
 
\section{Annex}
\begin{table}[H]
\let\center\empty
\let\endcenter\relax
\centering
\resizebox{\columnwidth}{!}{\begin{tabular}{llllllllllll}
\toprule
{} &     2000 &     2001 &     2002 &     2003 &     2004 &     2005 &     2006 &     2007 &     2008 &     2009 &     2010 \\
\midrule
n papers                    &    49,872 &    52,046 &    54,721 &    58,439 &    62,241 &    67,361 &    70,501 &    75,717 &    81,228 &    84,496 &    89,168 \\
mean of cited paper         &  27.3871 &  27.3672 &  27.9704 &  28.5654 &  28.8562 &  29.3572 &  30.0423 &  30.3297 &  31.0576 &  31.5393 &  32.3128 \\
var of cited paper          &  707.008 &  708.596 &  742.314 &  709.619 &  807.733 &  758.342 &  809.695 &  795.216 &   845.84 &  944.337 &  896.461 \\
mean of meshterms per paper &  13.3097 &  13.4067 &  13.2431 &  13.3788 &  13.2862 &  13.1364 &  12.8499 &  12.8425 &  12.8575 &  12.9128 &  12.8867 \\
var of meshterms per paper  &  26.5811 &  27.6517 &  26.0774 &  26.9045 &  27.2265 &  26.4599 &  22.9795 &  23.3933 &  23.3725 &  24.6855 &  24.9734 \\
\bottomrule
\end{tabular}
}
\caption{Sample Statistics}
\label{tab:table2}
\end{table}

\end{document}